\documentclass[prd,amsmath,amssymb,superscriptaddress,twocolumn,nofootinbib,10pt]{revtex4}

\usepackage{dcolumn}
\usepackage{bm}
\usepackage{amssymb}

\usepackage{booktabs}
\usepackage{amsmath}
\usepackage{multirow}
\usepackage{graphicx} 
\RequirePackage{xcolor}

\usepackage[colorlinks,linkcolor=red,citecolor=blue,anchorcolor=black,hyperfootnotes=true]{hyperref}
\newcommand{\beq}[1]{\begin{equation}\label{#1}}
 \newcommand{\eeq}{\end{equation}}
 \newcommand{\bea}{\begin{eqnarray}}
 \newcommand{\eea}{\end{eqnarray}}

\newcommand{\Mpc}{\mathrm{~km/s/Mpc}}

\newcommand\vertsp{\rule[-1.5mm]{1mm}{0mm} &}

\newcommand\morehorsp{\rule[-2.25mm]{0mm}{6mm}}

\def\({\left(}
\def\){\right)}
\def\[{\left[}
\def\]{\right]}

\def\aap{Astron. Astrophys.}
\def\mnras{Mon. Not. Roy. Astron. Soc.}
\def\apjl{Astrophys. J. Lett.}
\def\jcap{JCAP}
\def\nar{New Astronomy Reviews}
\def\apjs{Astrophys. J., Suppl. Ser.}
\def\pasp{Publications of the Astronomical Society of the Pacific}
\begin{document}

\title{Determination of cosmic curvature independent of the sound horizon and $H_0$ using BOSS/eBOSS and DESI DR1 BAO observations}

\author{Tonghua Liu}
\email{liutongh@yangtzeu.edu.cn}
\affiliation{School of Physics and Optoelectronic Engineering, Yangtze University, Jingzhou, 434023, China;}
\author{Shengjia Wang}
\affiliation{School of Physics and Optoelectronic Engineering, Yangtze University, Jingzhou, 434023, China;}
\author{Hengyu Wu}
\affiliation{School of Physics and Optoelectronic Engineering, Yangtze University, Jingzhou, 434023, China;}
\author{Jieci Wang}
\email{jcwang@hunnu.edu.cn}
\affiliation{Department of Physics, and Collaborative Innovation Center for Quantum Effects and Applications, Hunan Normal University, Changsha 410081, China;}

\begin{abstract}
We present an improved model-independent method for determining the cosmic curvature using the observations of Baryon Acoustic Oscillations (BAOs) and the Hubble parameter.   The purpose of this work is to provide insights into  late-universe  curvature measurements using available observational data and techniques. Thus, we use two sources of BAO data sets, BOSS/eBOSS and latest DESI DR1, and two reconstruction methods,  Gaussian process (GP) and artificial neural network (ANN).
It is important to highlight that our method circumvents influence induced by  the sound horizon in BAO observations  and the Hubble constant. Combining BAO data from BOSS/eBOSS plus DESI DR1, we find that the constraint on the cosmic curvature results in $\Omega_K=-0.040^{+0.142}_{-0.145}$ with an observational uncertainty of $1\sigma$ in the framework of GP method. This result changes to $\Omega_K=-0.010^{+0.405}_{-0.424}$ when the ANN method is applied.
Further comparative analysis of samples from two BAO data sources, we find that there is almost no difference between the two samples. Although the curvature values obtained from the data samples using DESI DR1 are on the slightly positive and the samples using BOSS/eBOSS are on the slightly negative, these results both report that our universe has a flat spatial curvature within uncertainties, and the precision of constraining the curvature with two BAO samples is almost equal.
\end{abstract}

\maketitle

\section{Introduction}           

\label{sect:intro}

The cosmic curvature  ($\Omega_K$) can be regarded as one of the most fundamental and crucial parameters in our universe, influencing its evolution and the properties of dark energy. Any deviation from a flat universe would prompt profound reflections and novel insights into inflation theory and fundamental physics. Recently, several studies have suggested that the spatial curvature of the universe remains an unresolved issue \citep{2021PhRvD.103d1301H,2021APh...13102605D}. More specifically, the combination of {\it Planck} lensing and low-redshift Baryon Acoustic Oscillations (BAOs) datasets has revealed a flat universe with a curvature of $\Omega_K=0.0007\pm0.0019$ \citep{2020A&A...641A...6P}. However, the recent work \citep{2020NatAs...4..196D} claims that pure {\it Planck} data supports a nearly closed universe ($\Omega_K=-0.044^{+0.018}_{-0.015}$)
, supported by Cosmic Microwave Background Radiation (CMBR) temperature and polarization power spectrum data.
It is important to emphasize that both methods rely, to some extent, on assumed specific cosmological models, namely the non-flat $\Lambda$ Cold Dark Matter ($\Lambda$CDM) model. This tension, akin to the Hubble constant tension (for further discussion on the Hubble tension, please refer to references \cite{2019PhRvL.122v1301P,2019PhRvL.122f1105F,2021ApJ...912..150D,2022A&A...668A..51L,2020ApJ...895L..29L,
2023arXiv230306974D,2023ApJS..264...46L,2022Galax..10...24D}), is based on the $\Lambda$CDM model. Therefore, curvature measurements independent of cosmological models can also serve as a test for the $\Lambda$CDM model and are worth pursuing. Similarly, any curvature tension related to the CMB may be another piece of evidence against the $\Lambda$CDM model.

All these crises may stem from inconsistencies in measurements between the early and late universes, as pointed out in numerous works \citep{2023PhRvD.108f3522Q,2021ApJ...908...84V,2022JHEAp..33...10Z,2022PhRvD.105f3524D}, which highlight that relying on the $\Lambda$CDM model can lead to such discrepancies. To address the crisis posed by measurements of cosmic curvature, a necessary approach is to confirm the late-universe value of $\rm\Omega_K$ using a method independent of cosmological models. A non-exhaustive references on this topic can be found in \citep{2019PhRvL.123w1101C,2020ApJ...897..127W,2021MNRAS.503.2179Q,2022ApJ...939...37L,
2023PhRvD.107f3522D,2023MNRAS.523.3406F,2021MNRAS.506L...1D}. A typical approach for curvature measurement involves utilizing the latest observational data provided by high-precision cosmological probes, namely the distance modulus data of Type Ia supernovae (SNe Ia), radial BAO measurements from cosmic chronometers and Hubble parameters, to obtain constraints on the geometric quantity $\Omega_K$. However, it should be noted that BAO data are not entirely independent of cosmological models, as we need to assume a prior value for the sound horizon $r_d$ inferred from CMB observations. In other words, the prior value of $r_d$ obtained from CMB analysis is used to calibrate BAO distances. Furthermore, previous works generally employed non-parametric Gaussian Process (GP) methods to reconstruct the expansion history of the universe. However, as mentioned in numerous  works \citep{2018MNRAS.478.3640M,2022JCAP...12..029M}, one of the most debated topics recently concerning the non-parametric technique of GP reconstruction in cosmology is that it reveals some fundamental issues, such as over-fitting and kernel consistency problems \cite{2021arXiv210108565C}. Currently considering the rapid development of algorithms, many machine learning methods such as artificial neural network (ANN) methods are applied in astronomy and cosmology. Therefore, along with different machine learning reconstruction methods and thus robustness conclusions are still worth exploring.

Inspired by the aforementioned considerations, we propose an improved method independent of cosmological models to measure cosmic curvature by combining BAO line-of-sight and transverse observational data with cosmic chronometer (CC) datasets using different machine learning methods. The primary advantage of this work is that it eliminates the need for a prior value of the sound horizon ${r_d}$ to perform cosmology model-independent curvature measurements.  The organization of this paper is as follows: In Sec. 2, we briefly introduce the methods and observational data used in our work. Two machine learning reconstructed methods and results are given in Sec. 3. Finally, we summarize and make discussions in Sec. 4.

\section{Methodology}\label{sec2}
\subsection{BAO data from the BOSS/eBOSS, and DESI 2024 surveys}
\begin{table*}
	\centering
	\begin{tabular}{ccccc}
		\toprule
		Data source \vertsp $z$ \vertsp $D_M/r_d$ \vertsp $D_H/r_d$  \vertsp Reference  \\
		\hline\hline
		\morehorsp
		BOSS galaxy--galaxy \vertsp $ 0.38 $ \vertsp $ 10.23\pm 0.17 $ \vertsp $ 25.00\pm 0.76 $ \vertsp \cite{2017MNRAS.470.2617A} \\
		eBOSS galaxy--galaxy \vertsp $ 0.51 $ \vertsp $ 13.36\pm0.21 $ \vertsp $ 22.33 \pm 0.58 $ \vertsp \cite{2017MNRAS.470.2617A}\\
		\vertsp $ 0.70$ \vertsp $ 17.86 \pm 0.33 $ \vertsp $ 19.33\pm 0.53 $ \vertsp \cite{2021MNRAS.500..736B,2020MNRAS.498.2492G}\\
		\vertsp $ 0.85 $ \vertsp $ 19.50\pm 1.00 $ \vertsp $ 19.60\pm 2.10 $ \vertsp \cite{2020MNRAS.499.5527T,2021MNRAS.501.5616D}\\
		\vertsp $1.48$ \vertsp $ 30.69\pm0.80 $ \vertsp $ 13.26\pm 0.55 $ \vertsp \cite{2020MNRAS.499..210N,2021MNRAS.500.1201H}\\
		eBOSS Ly-$\alpha$--Ly-$\alpha$ \vertsp $ 2.33 $ \vertsp $ 37.50 \pm 1.10 $ \vertsp $ 8.99\pm0.19 $ \vertsp \cite{eBAO2} \\
		eBOSS Ly-$\alpha$--quasar \vertsp $ 2.34 $ \vertsp $ 37.41\pm1.86 $ \vertsp $ 8.86\pm0.29 $
\vertsp \cite{eBAO1} \\
\hline
\hline
\morehorsp
		DESI DR1 LRG \vertsp $ 0.51 $ \vertsp $ 13.62\pm 0.25 $ \vertsp $ 20.98 \pm 0.61 $
\vertsp \cite{2024arXiv240403002D} \\
		DESI DR1 LRG \vertsp $ 0.71 $ \vertsp $16.85 \pm0.32 $ \vertsp $ 20.08 \pm 0.60 $
\vertsp \cite{2024arXiv240403002D} \\
		DESI DR1 LRG+ELG \vertsp $ 0.93 $ \vertsp $21.71 \pm 0.28 $ \vertsp $ 17.88 \pm 0.35 $
\vertsp \cite{2024arXiv240403002D} \\
		DESI DR1 ELG \vertsp $ 1.32 $ \vertsp $27.79 \pm 0.69 $ \vertsp $ 13.82 \pm 0.42 $
\vertsp \cite{2024arXiv240403002D} \\
		DESI DR1 Ly-$\alpha$--quasar \vertsp $ 2.33 $ \vertsp $39.71\pm 0.94 $ \vertsp $ 8.52 \pm 0.17 $
\vertsp \cite{2024arXiv240403002D} \\
		\hline\hline
	\end{tabular}
	\caption{ The BAO data used in this work given in terms of $D_M = (1+z)D_A$ and $ D_H = c\,H(z)^{-1}$.}
	\label{tab1}
\end{table*}

In the early universe, BAOs originated from the intricate gravitational interactions between the photon-baryon fluid and inhomogeneities, a mechanism first elaborated in detail by \citet{1972gcpa.book.....W}. As the universe evolved and entered the drag epoch, baryons gradually decoupled from photons, ultimately freezing out at a scale corresponding to the sound horizon at the drag redshift. This specific scale, known as the sound horizon scale, serves as an intrinsic standard for galaxy distribution and is one of the crucial predictions of cosmological theoretical models. Its size depends on the speed of sound in the baryon-photon plasma and the expansion rate of the early universe prior to decoupling of matter and radiation.  Sloan Digital Sky Survey (SDSS) \cite{2005ApJ...633..560E} and the 2dF Galaxy Redshift Survey (2dFGRS) \cite{2005MNRAS.362..505C} pioneered the detection of BAO signals, powerfully validating the remarkable value of BAOs as a cosmological standard ruler. Following this, a series of BAO surveys aimed at enhancing the precision of cosmological distance measurements were launched, including the Six-degree Field Galaxy Survey (6dFGS) \cite{2011MNRAS.416.3017B}, the Baryon Oscillation Spectroscopic Survey (BOSS) \cite{2017MNRAS.470.2617A}, the extended Baryon Oscillation Spectroscopic Survey (eBOSS) \cite{2021PhRvD.103h3533A}, and the WiggleZ Survey \cite{2012MNRAS.425..405B}.  These measurements could provide a statistically significant insight into the ongoing tension problems in cosmology (for a more comprehensive discussion, please refer to \cite{2021CQGra..38o3001D,2022JHEAp..34...49A,2022NewAR..9501659P} and the references cited therein).

Currently, the Dark Energy Spectroscopic Instrument (DESI) collaboration has released Data Release 1 (DR1) of BAO measurements, presenting evidence for dynamical dark energy at a significance level exceeding $2\sigma$ \cite{2024arXiv240403000D,2024arXiv240403001D,2024arXiv240403002D}. DESI covers six
different classes of tracers, including low redshift galaxies of the bright galaxy survey (BGS),
luminous red galaxies (LRG), emission line galaxies (ELG), quasars as direct tracers, and
Lyman-$\alpha$ (Ly-$\alpha$) forest quasars to trace the distribution of neutral hydrogen.  DESI is designed to  achieve more accurate percent-level measurements of cosmological distances, paving the way for uncovering the deeper mysteries of the universe.

The observations of BAO provide line-of-sight and transverse measurements, but these entangle with the radius of the sound horizon $r_d$, i.e., $D_M/r_d$ and $D_H/r_d$. The terms
\begin{equation}
D_M=(1+z)D_A,\,\,\,D_H=cH(z)^{-1},
\end{equation}
is transverse and line-of-sight directions measurement in BAO observations.
Before employing BAO as a standard ruler in cosmology and leveraging it as a potent probe of the universe, it is imperative to ascertain the comoving length of this ruler, i.e., the sound horizon $r_d$ at the epoch of radiation drag. It is common procedure to use the prior value of radius of the sound horizon $r_d$ obtained from \textit{Planck} CMB observations. However, it is to some extent cosmological model-dependent, i.e., partly dependent on the assumptions of cosmological models. Taking this into account, we use the BAO measurements in this work and eliminate this effect by our new method.

We only need to combine the BAO observations in the transverse and line-of-sight directions with additional $[H(z)]^{CC}$ provided by the CC dataset to realize the angular diameter distances  measurements independent of the sound horizon $r_d$ and cosmological models, which is given by
\begin{equation}\label{BAODA}
[D_A(z)]^{BAO+CC}=\frac{c}{(1+z)\,[H(z)]^{CC}}\cdot\bigg[\frac{D_M}{r_d}\cdot\frac{r_d}{D_H}\bigg]^{BAO}.
\end{equation}
From the equation above, it can be inferred that if we know the Hubble parameter $H(z)$, then, in conjunction with observations of BAO, we can derive the angular diameter distance (absolute distance), and this can be utilized to determine the curvature parameter.
For the BAO data, we consider two sources of data, i.e., seven data points from BOSS/eBOSS, and five data points from DESI 2024 surveys. These observed data for BAO given in Table \ref{tab1}.

\subsection{Reconstruction for comoving distances from cosmic chronometer with Gaussian process and Artificial Neural Network}
The cosmic chronometers (CCs), serving as the standard clocks, offer measurements of the Hubble parameter. These measurements are derived from the integration of differential age estimations of systems characterized by passively evolving star populations, such as globular clusters, with their corresponding spectral redshifts \cite{2002ApJ...573...37J},
\begin{equation}
[H(z)]^{  CC}=-\frac{1}{1+z}\frac{\Delta z}{\Delta t}.
\end{equation}
The CC data are obtained by measuring the ages of red-envelope galaxies using different age methodologies. The aging of stars can be regarded as an indicator of cosmic aging. The spectra of stars can be converted into information about their ages, given that the evolution of stars is well-understood. Since stars cannot be observed individually on a cosmic scale, the spectra of relatively homogeneous populations of stars within galaxies are typically used. This method relies on the detailed shape of the galaxy spectrum, rather than the luminosity of the galaxy. Consequently, CC data do not require calibration \cite{2010JCAP...10..024A,2015PhRvD..92l3539L,2021PhRvL.126w1101K}.
The comprehensive details of the 32 CCs are presented in Table 1 of the literature. The redshift range covered by these 32 CC data spans from 0.07 to 1.965 \cite{2014RAA....14.1221Z,2010JCAP...02..008S,2012JCAP...08..006M,2016JCAP...05..014M,2017MNRAS.467.3239R,2015MNRAS.450L..16M}.
To achieve cosmological model-independent and the sound horizon-independent measurements of curvature,  it is necessary to reconstruct a smooth curve of $[H(z)]^{  CC}$ with a non-parametric reconstruction technique firstly. We consider two machine learning techniques to reconstruct the Hubble parameter, which are served for the BAO data.

\begin{figure*}
\includegraphics[scale=0.35]{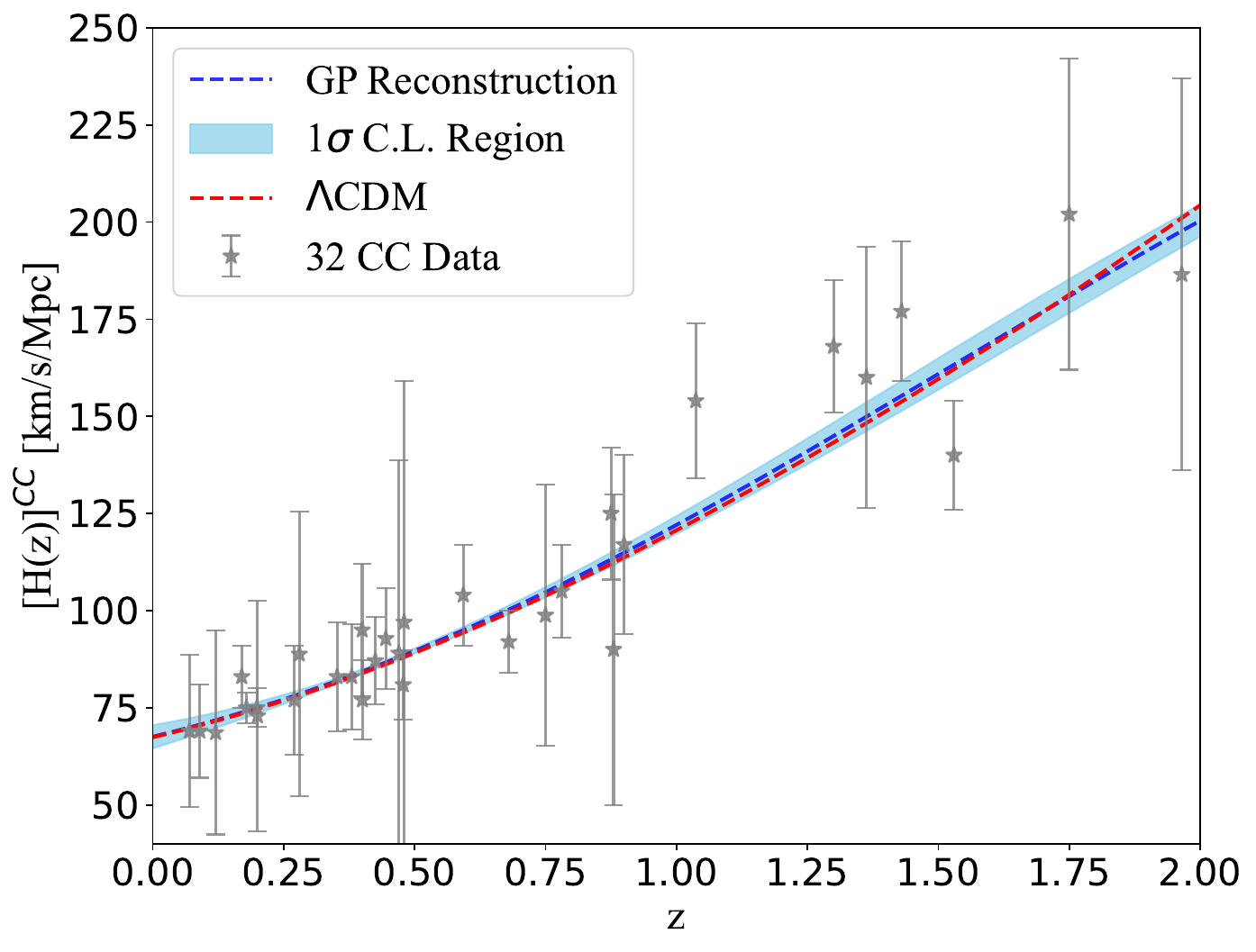}
\includegraphics[scale=0.35]{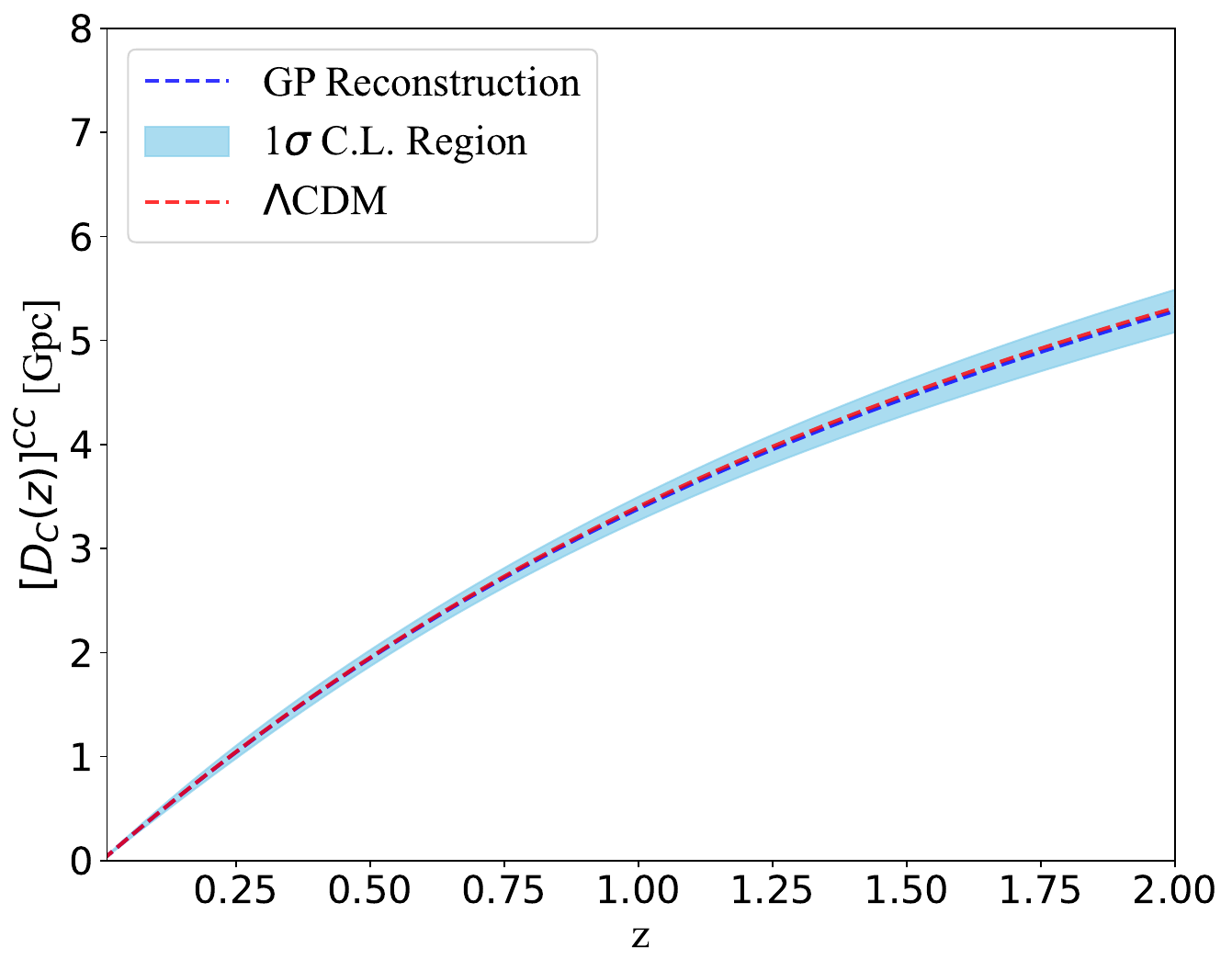}
\caption{Left: The reconstructions of $[H(z)]^{CC}$ from CC data by using GP method. The blue region and the blue solid line denote the 1$\sigma$ confidence level uncertainty and the best-fit values of reconstruction. The gray points with error bars represent the observed CC data. Right: The  comoving distance  $[D_C(z)]^{CC}$  from the  reconstructed $[H(z)]^{CC}$ by using integrating. The red dotted line is the fiducial flat $\Lambda$CDM model with $\Omega_m=0.30$ and $H_0=70.0 \Mpc$.}
\label{fig:reconstruction}
\end{figure*}

\textit{Gaussian Process.---} For the first reconstruction method, we choose the Gaussian process. We provide a concise overview of the GP methodology, which facilitates the straightforward reconstruction of a function from data without necessitating any parametric assumptions. We utilize the widely-adopted \texttt{GaPP} Python package, prevalent in cosmological applications, to implement the GP method. Within this framework, we assume that the values of the reconstructed function  $f(z)$ at two distinct points$z$ and $\tilde{z}$
are correlated through a covariance function $k(z,\tilde{z})$, which solely depends on two hyperparameters: $\sigma_f$ and $\ell$. Despite the existence of numerous and effective covariance function forms, according to the analysis conducted by \citet{2013arXiv1311.6678S}, the squared exponential form, specifically with theMat\'{e}rn $(\nu=9/2)$  covariance function, yields more reliable results compared to all others. Therefore, we adopt this form here, and its expression is given by:
\begin{eqnarray}
k(z,\tilde{z})&=&\sigma_f^2\exp(-\frac{3|z-\tilde{z}|}{\ell})\times (1+\frac{3|z-\tilde{z}|}{\ell} \nonumber\\
&+&\frac{27(z-\tilde{z})^2}{7\ell^2}
+\frac{18|z-\tilde{z}|^3}{7\ell^3}+\frac{27(z-\tilde{z})^4}{35\ell^4}).\label{7}
\end{eqnarray}
where  the hyperparameter $\ell$  represents the characteristic length scale, indicating the distance over which the function $f(z)$ undergoes significant changes. The hyperparameter $\sigma_f$ denotes the typical variation or dispersion in the observed data. These hyperparameters, $\sigma_f$ and $\ell$, are marginalized. Notably, the optimization of these two hyperparameters is conducted independently of the fitting process for cosmological parameters. A more detailed description of these hyperparameters, the kernel, and the GP method can be found in the literature \cite{2012PhRvD..85l3530S,2023JCAP...02..014H,2019ApJ...886L..23L,2024MNRAS.528.1354L,2024ApJ...960..103L}.
The final reconstructed $[H(z)]^{  CC}$  from the CC dataset are shown in the left panel of the Fig.~ \ref{fig:reconstruction}.

In the framework of the Friedmann-Lematre-Robertson-Walker (FLRW) metric, the comoving distance $[D_C(z)]^{  CC}$ is defined as
\begin{eqnarray}
[D_C(z)]^{CC}=c\int_{0}^{z}\frac{d z^{\prime}}{[H(z^{\prime})]^{  CC}}\label{con:E2.4},
\end{eqnarray}
where $c$ is the speed of light. Using the reconstructed smooth function $[H(z)]^{CC}$, we integrate this function to obtain the smooth function $[D_C(z)]^{CC}$. By integrating the errors of $[H(z)]^{CC}$, we derive the confidence regions for the smooth function. The comoving distance $[D_C(z)]^{CC}$ is displayed in the right panel of Fig. \ref{fig:reconstruction}.

\begin{figure*}
\includegraphics[scale=0.35]{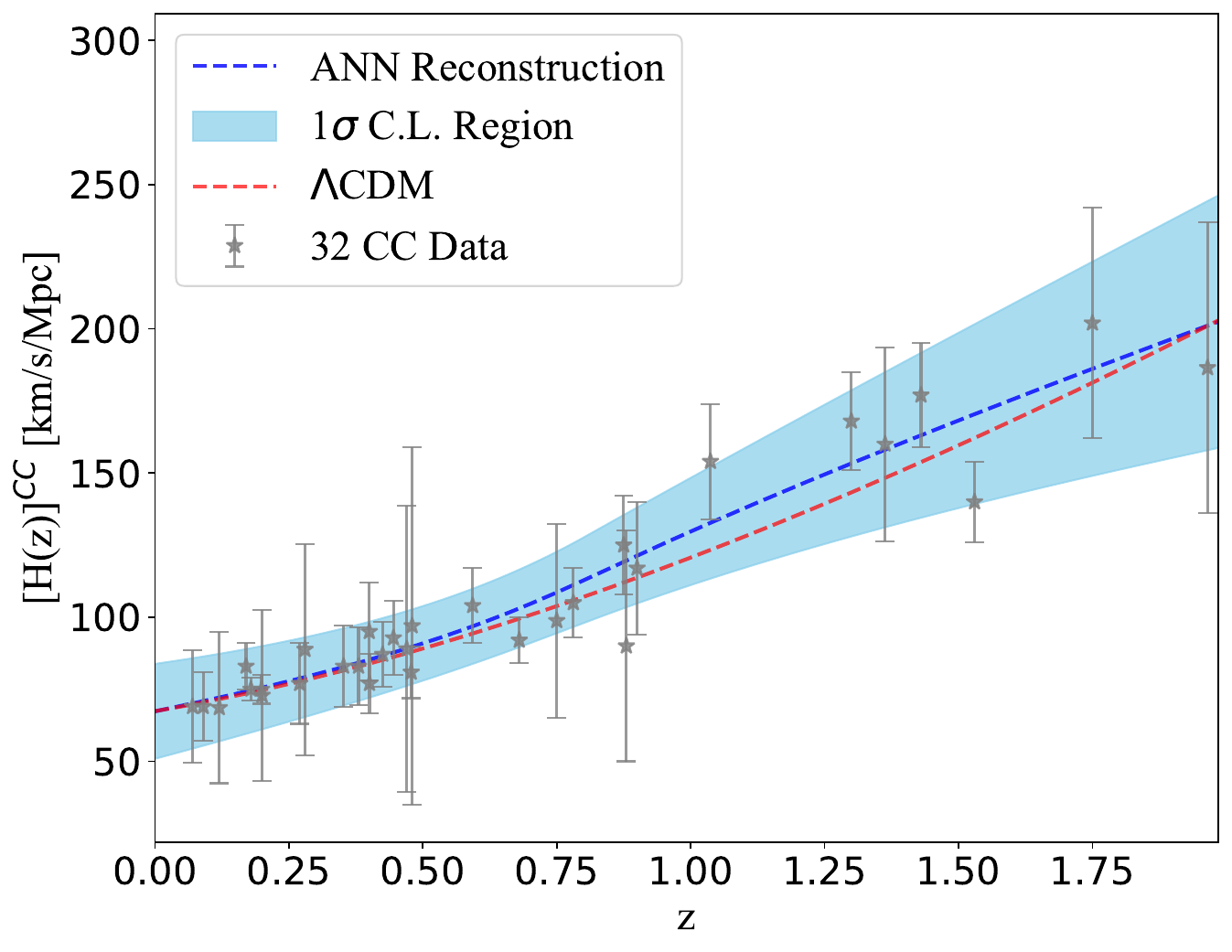}
\includegraphics[scale=0.35]{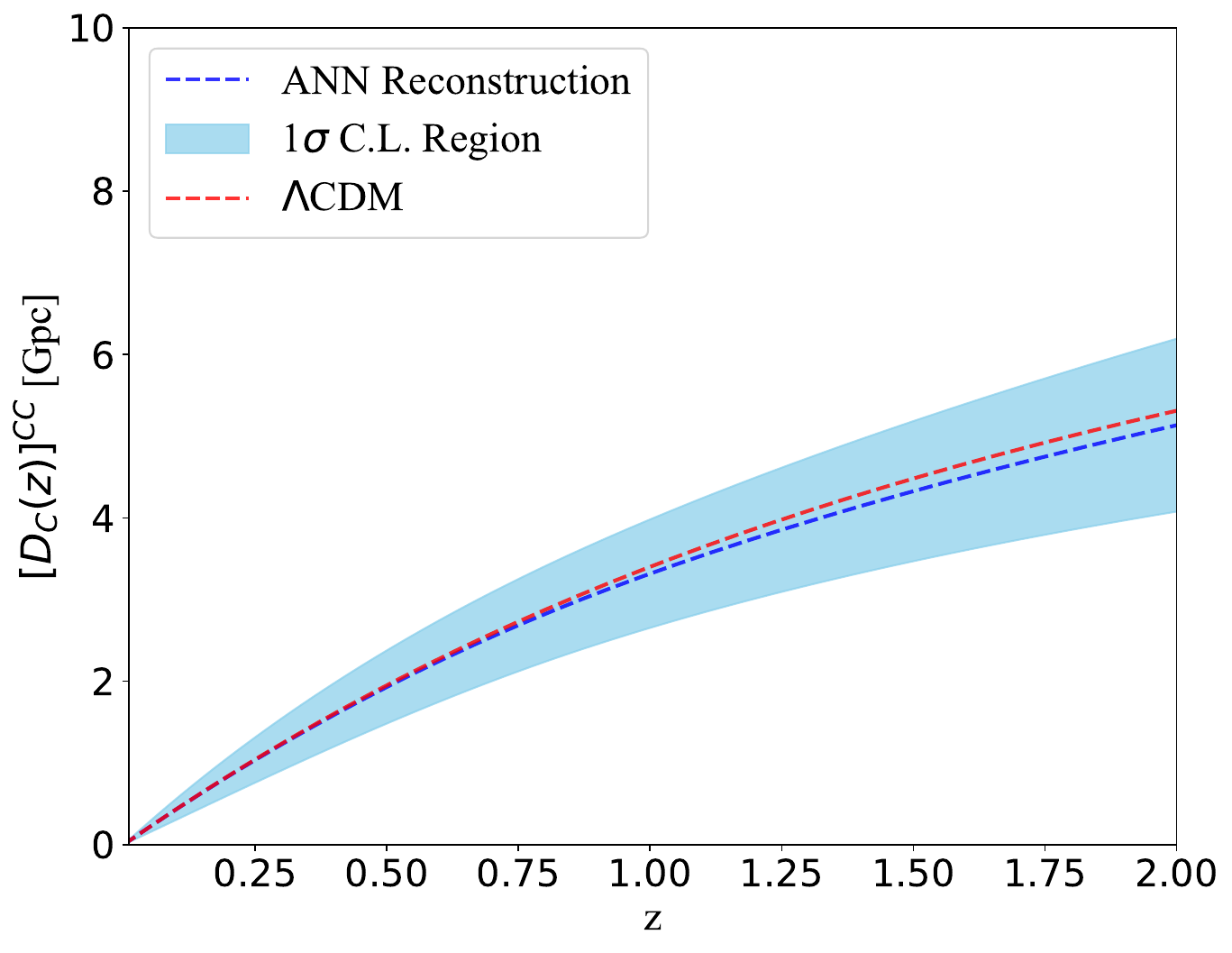}
\caption{Same as the Fig. 1 but using ANN method.}
\label{fig:ANN}
\end{figure*}

\textit{Artificial Neural Network.---}For the second reconstruction method, we choose to utilize the ANN approach.
Here, we employ the ANN method based on the \texttt{REFANN}\footnote{https://github.com/Guo-Jian-Wang/refann} Python code \citep{2020ApJS..246...13W} to reconstruct the  $[H(z)]^{  CC}$ function from the data. This method has also been widely applied in cosmology \citep{2024PDU....4601706M,2024ApJ...965L..11L,2024PhRvD.109d3001R,2023PhRvD.108f3522Q,2023EPJC...83..304G}.
The ANN method is a non-parametric approach that does not assume Gaussian-distributed random variables and is fully data-driven. The optimal ANN model used for our reconstruction function is the same as the one selected in \citet{2020ApJS..246...13W}, which features a single hidden layer with a total of 4096 neurons. The $[H(z)]^{  CC}$ function reconstructed using the ANN is also shown in the left panel of Fig. \ref{fig:ANN}. The final comoving distance $[D_C(z)]^{  CC}$, obtained by integrating the $[H(z)]^{  CC}$ function, is displayed in the right panel of Fig. \ref{fig:ANN}.

From the reconstruction results, it can be observed that the uncertainties in the data reconstructed by the ANN are generally equivalent to the uncertainties in the observed values. Meanwhile, the confidence regions reconstructed by the ANN are larger than those reconstructed by the GP method, which may be related to the fundamental logic and nature of the two methods. GP is a probabilistic modeling approach that infers unknown functions by defining probability distributions over function spaces. It focuses on reconstructing smooth functions based on the covariance between data points, emphasizing the overall structure and correlation of the data. In contrast, ANN is computational modeling method that processes data by simulating the structure and function of biological neural networks. More   Secondly, the training processes differ significantly between GP and ANN methods. The training of GP primarily relies on the parameter estimation of covariance functions, often using methods such as maximum likelihood estimation or Bayesian optimization. Since GP is probabilistic modeling methods, the probability distribution and uncertainty of the data need to be considered during the training process. In contrast, ANN typically uses the back-propagation algorithm during training, adjusting weights and biases through optimization methods such as gradient descent. During the training process of ANN, the choice of loss function, adjustment of learning rate, and application of regularization techniques are crucial in preventing over-fitting. Please refer to refs \cite{2013arXiv1311.6678S} for the GP method and  \cite{2020ApJS..246...13W} for the ANN method with more details.

Furthermore, it is important to acknowledge that GP and ANN have distinct advantages and limitations. GP excels at capturing uncertainty and estimating smooth functions, while ANN is highly effective in modeling complex nonlinear relationships. These inherent differences in modeling techniques may lead to variations in how the methods handle noise, outliers, and subtle features in the data, resulting in different reconstructions of $[H(z)]^{  CC}$.
The results of our reconstructing are almost same as the previous work \cite{2023PhRvD.108f3522Q,2021EPJC...81..903L}, and their work also demonstrates the differences between the two reconstructing approaches.
Finally, it should be noted that  we have not extrapolated beyond the redshift range. Considering the uncertainty of the reconstructed data at the extrapolated redshift, we strictly limit the range of our reconstruction function to the maximum redshift range of $[H(z)]^{ CC}$, which means that the two BAOs data points from BOSS/eBOSS surveys will be discarded, as well as the one  BAO data points from DESI survey.

\subsection{Methodology for directly measuring cosmic curvature }
The cornerstone of modern cosmology rests upon the fundamental principles of the universe: our universe exhibits homogeneity and isotropy on a large scale. The space-time of our universe can be depicted using the FLRW metric:
\begin{equation}\label{eq1}
ds^2=dt^2-\frac{a(t)^2}{1-Kr^2}dr^2-a(t)^2r^2d\Omega^2,
\end{equation}
where $a(t)$ denotes the scale factor, and $K$ represents a dimensionless curvature parameter that adopts one of three values: $\{-1, 0, 1\}$, corresponding to a closed, flat, or open universe, respectively. The cosmic curvature parameter, denoted as $ \Omega_K$, is related to $K$ and the Hubble constant
 $H_0$ through the  $ {\Omega_K}$ $=-c^2K/a^2_0H_0^2$.
The angular diameter distance, $D_A$, and the comoving distance, $D_C$ , satisfy the following relationship \cite{1972gcpa.book.....W}:
\begin{equation}\label{eq3}
{D_A}(z;\Omega_K) = \left\lbrace \begin{array}{lll}
\frac{c}{(1+z)H_0\sqrt{ \Omega_{  K}}}\sinh\left[\sqrt{ \Omega_{  K}}{\frac{H_0}{c}D_C}\right]~~~ \Omega_{K}>0,\\
	\morehorsp
D_C/(1+z)~~~~~~~~~~~~~~~~~~~~~~~~~~~~ \Omega_{K}=0, \\
\frac{c}{(1+z)H_0\sqrt{| \Omega_{  K}|}}\sin\left[\sqrt{| \Omega_{  K}|}\frac{H_0}{c}D_C\right] \Omega_{K}<0.\\
\end{array} \right.
\end{equation}
As can be seen from the Eq. (\ref{eq3}), to measure the curvature in a cosmological model in a manner that is independent of other assumptions, it is necessary to obtain both the comoving distance and the angular diameter distance at the same redshift.
We determine the cosmic curvature by maximizing  likelihood function $\mathcal{L}$:
\begin{equation} \label{Eq:chi2}
\ln (\mathcal{L})=-\frac{1}{2}\Delta D_i^T {\textbf{Cov}_{ij}}^{-1} \Delta D_j,
\end{equation}
where $\Delta D$ is the  angular diameter distance residuals computed as $\Delta D=[D_A(z_i)]^{  BAO+CC}-[D_A(z_i;\Omega_K)]^{  CC}$ from the Eqs. (\ref{BAODA}), and (\ref{con:E2.4}) embedded into (\ref{eq3}).
The covariance matrix consists of two parts, $\textbf{Cov}_{ij}=\textbf{Cov}_{ii}^{\text{stat}}+\textbf{Cov}_{ij}^{\text{corr}}$, where the diagonal elements $\textbf{Cov}_{ii}^{\text{stat}}=\sigma_{[D_A(z_i)]^{  BAO+CC}}^2+\sigma_{[D_A(z_i)]^{  CC}}^2$ contain the reconstruction error of $[H(z)]^{  CC}$ plus the observation error of BAO, and the reconstruction error of $[D_C(z)]^{  CC}$, and both are passed through the standard error propagation formula to angular diameter distance.
It is worth noting that the non-diagonal element part, $\textbf{Cov}_{ij}^{\text{corr}}$, due to the fact that $[D_A(z)]^{  CC}$ is derived from the integral of $[H(z)]^{  CC}$, while the data of $[H(z)]^{  CC}$ itself is used simultaneously in the combination of the $[D_A(z)]^{  BAO+CC}$  data, so we additionally consider the  covariance between the $[D_A(z)]^{ BAO+CC}$ and $[D_A(z)]^{CC}$. The calculation of the  covariance  matrix is based on the standard definition of this matrix, i.e., $\textbf{Cov}_{ij}^{\text{corr}}=Cov(X_i,Y_j)=E[(X_i-\mu_i)(X_j-\mu_j)]$.

Finally, we constrain the cosmic curvature $\Omega_K$ using the \texttt{emcee} Python module based on the Markov Chain Monte Carlo analysis \citep{2013PASP..125..306F}. To perform the above  procedures, we have to take an a priori value for the Hubble constant, and a natural way to do this is to take the reconstructed value of $H_0=H(z=0)$ as a prior for the Hubble constant.
Let us emphasize here that after our tests we have found that the value of $H_0$ does not have any effect on our whole work. From a theoretical point of view, the Hubble constant affects both $[D_A(z)]^{  BAO+CC}$ and $[D_A(z)]^{  CC}$, so changing the value of $H_0$ does not affect the determination of curvature.

\section{Results and discussion} \label{sec:method}

\begin{figure*}
{\includegraphics[scale=0.4]{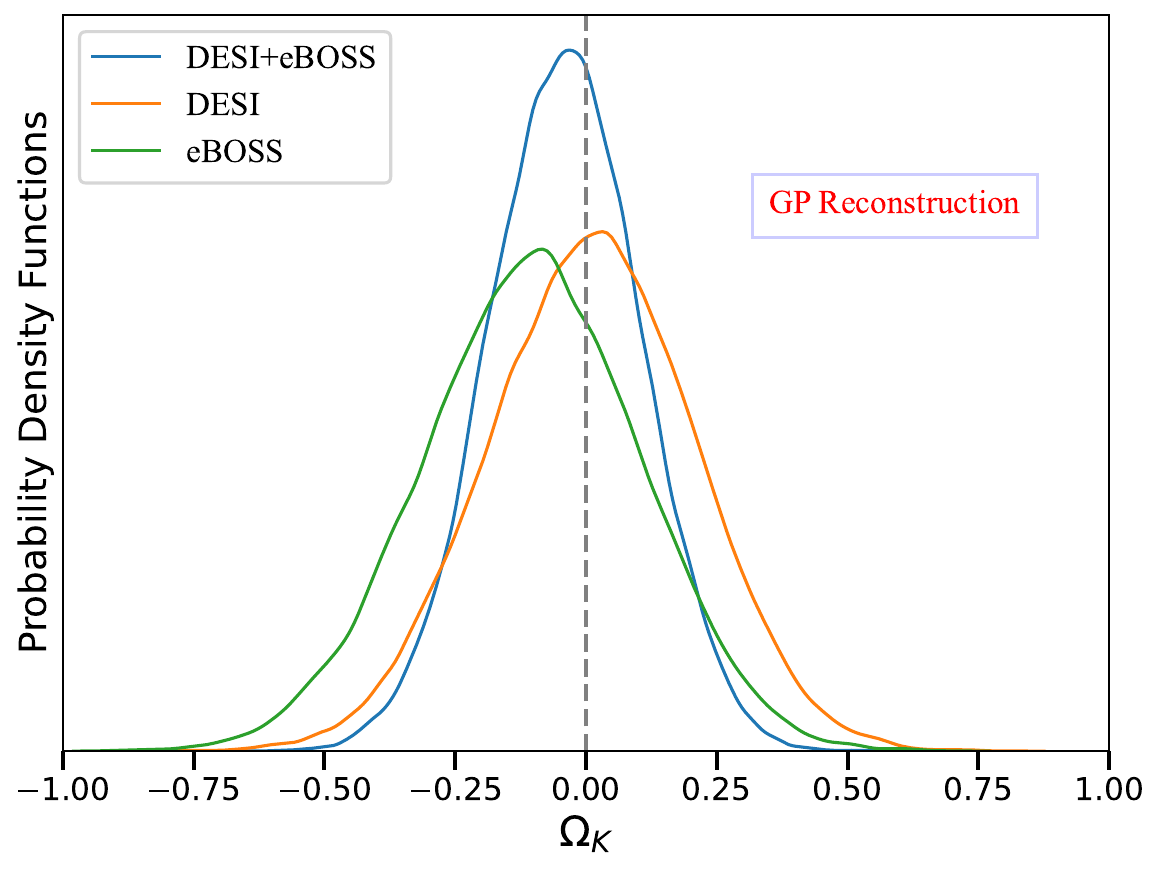}
\includegraphics[scale=0.4]{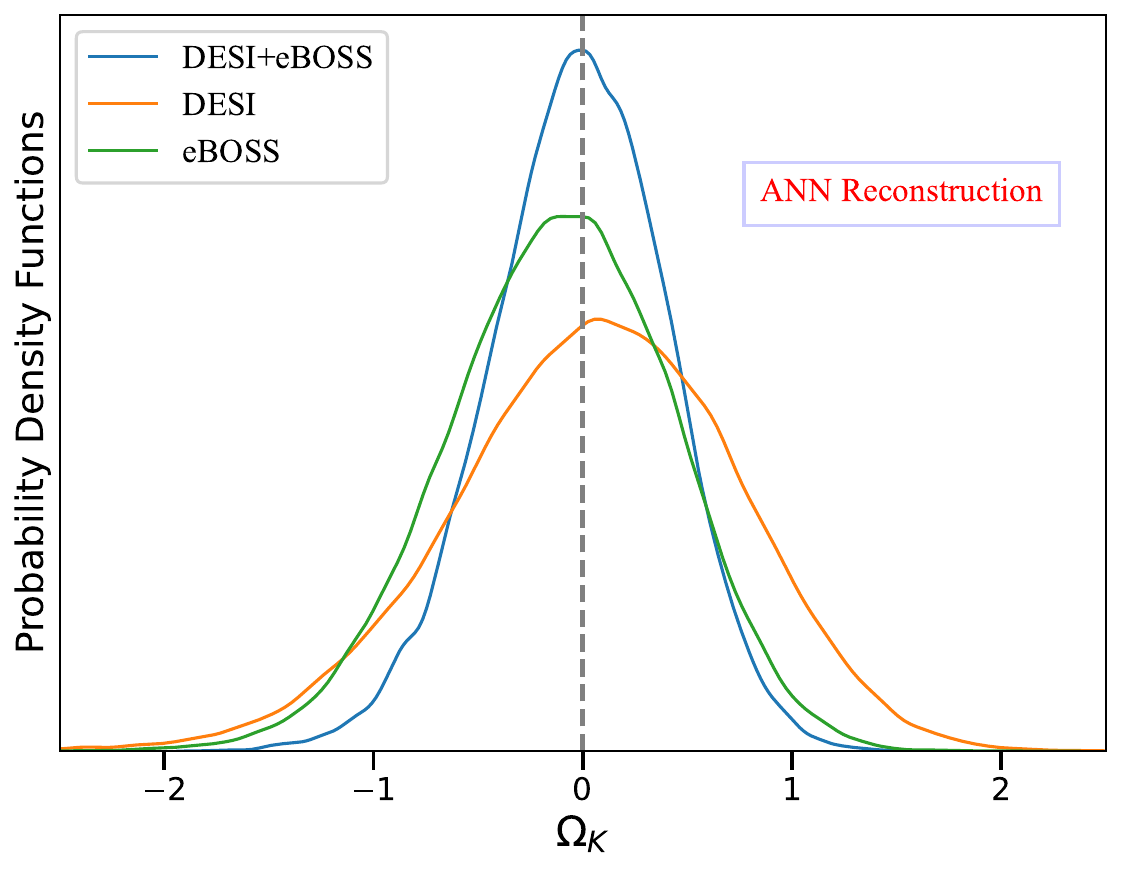}}
\caption{\textit{Left panel}: The measurements on $\Omega_K$ using the GP method. \textit{Right panel}: The measurements on $\Omega_K$ using the ANN method.}\label{fig3}
\end{figure*}

Let's emphasize the advantages of our work. First, the determination of the curvature does not require any cosmological model, all the quantities involved come from observations; second, it does not involve any nuisance parameters, in particular the Hubble constant and the  sound horizon parameter in the data from the BAO; and third, it does not require assumptions on the Eddington relation, i.e., the distance duality relation relation.

Regarding the GP method and working on the DESI DR1 BAO data, we obtain the best-fitting value $ \Omega_K =0.012^{+0.194}_{-0.176}$ with 1$\sigma$ confidence level. The best-fitting value for cosmic curvature is positive, and  support a flat universe at $1\sigma$ confidence level.
Combined with BOSS/eBOSS  BAO data, a flat universe is also supported within the 1$\sigma$ uncertainty, but the center value is a little negative. In addition, we find an interesting finding that using data from different sources of BAO does not significantly affect the precision of the constrained $ \Omega_K$.
Combined with BOSS/eBOSS plus DESI DR1 BAO data,  we obtain the best-fitting values are $ \Omega_K=-0.040^{+0.142}_{-0.145}$ with 1$\sigma$ uncertainty. This result is consistent with the  value inferred from Planck data \cite{2020A&A...641A...6P}, i.e., a flat universe is  supported within the 1$\sigma$ uncertainties, but the center value is a little negative.
These posterior distributions of the $ \Omega_K$  are shown in the left panel of the Fig. \ref{fig3}. The  numerical results of the constraints on the $ \Omega_K$ are shown in Table \ref{tab2}.

For the ANN method, we find that the precision of the $ \Omega_K$ obtained through this method is inferior to that achieved by the GP method. This discrepancy can be traced back to results of two different reconstructed $H(z)$ functions, as previously mentioned, the confidence region for $H(z)$  reconstructed using the ANN is significantly broader compared to the confidence region reconstructed with the GP method. Nevertheless, However, the center values of the curvature parameters obtained by both methods are almost the same. This indirectly validates the robustness of our approach, indicating that different data reconstruction methods do not significantly impact the accurate value of the curvature parameter. Meanwhile, we also obtain a conclusion that, throughout the entire process of constraining curvature, the uncertainty level of $H(z)$ remains the primary source of error compared with uncertainty of BAO.
The posterior distributions of the $ \Omega_K$ using the ANN method are shown in the right panel of the Fig. (\ref{fig3}), and  numerical results of the constraints on the $ \Omega_K$ are given in Table \ref{tab2}.

\begin{table}
\renewcommand\arraystretch{2.2}
\caption{\label{tab2} Summary of the constraints on the spatial curvature parameter $ \Omega_K$ by using different data combination and different reconstructed methods.}

\begin{centering}
\renewcommand{\arraystretch}{2.2}
\begin{tabular}{c| c| c }
\hline
\hline
Data Combination  & GP method ($ \Omega_K$) &ANN method ($ \Omega_K$) 
\\
\hline
 DESI   & $0.012^{+0.194}_{-0.176}$ & $0.075^{+0.649}_{-0.692}$ \\
\hline
 eBOSS  & $-0.103^{+ 0.205}_{-0.209}$ & $-0.100^{+0.520}_{-0.551}$ \\
\hline
 DESI+ eBOSS   & $-0.040^{+0.142}_{-0.145}$ & $-0.010^{+0.405}_{-0.427}$ \\
\hline
\hline
\end{tabular}
\end{centering}
\end{table}

Recent studies have suggested that the $H_0$ tension might be attributed to inconsistencies in spatial curvature between the early and late universes \cite{2020NatAs...4..196D,2021PhRvD.103d1301H,2021APh...13102605D}. An effective approach to investigating this issue involves simultaneously constraining both the cosmic curvature and the Hubble constant, employing methods that are independent of cosmological models \citep{2019PhRvL.123w1101C,2020ApJ...897..127W,2021MNRAS.503.2179Q,2022ApJ...939...37L}. This is due to the degeneracy that exists between cosmic curvature and the Hubble constant.
However, our method completely avoids the presence of this degeneracy, as the entire process is independent of both $H_0$ and  the sound horizon. Although the precision of our constrained results on $\Omega_K$ is not exceptionally high compared to previous work, it is reasonable given that we only utilized BAO data from nine points (four from DESI DR1 and five from eBOSS). From a statistical perspective, when the number of BAO observations increases by a factor of $N$, the precision should improve by a factor of $N$. Regardless of the reconstruction method employed, our results demonstrate this point: the precision of the curvature obtained using either DESI or eBOSS data alone is $1/\sqrt{2}$ times that obtained using the combined DESI + eBOSS data. Taking this aspect into consideration, we look forward to a large amount of future data, not only from the observations of BAO, but also from the cosmic chronometer, allowing us to further improve the precision of $\Omega_K$ measurements.

\section{Conclusion}
In this paper, we present an improved model-independent method for determining the cosmic curvature using BOSS/eBOSS and DESI DR1 measurements of BAOs and observations of the Hubble parameter.  All quantities involved are derived from observations. Importantly, our method circumvents influence induced by  the sound horizon in BAO observations  and the Hubble constant. The objective of this work is to conduct a thorough investigation of late-universe measurements using our existing observational data and statistical tools. Therefore, we consider two sources of BAO data sets, BOSS/eBOSS and latest DESI DR1, and two reconstruction methods, GP method and ANN method.

The final result obtained combining all BAO measurements gives $\Omega_K=-0.040^{+0.142}_{-0.145}$ with $1\sigma$ confidence level in the framework of GP method. This result changes to  $\Omega_K=-0.010^{+0.405}_{-0.424}$ when the ANN method is used.  Note that the GP method has yielded the most precise constraint on $\Omega_K$, with a constraint precision of $\Delta \Omega_k=0.14$  for the combined set of all BAO data. This surpasses recent measurements of $\Omega_K$ using similar methods \citep{2021MNRAS.501.5714W,2017ApJ...838..160W,2016ApJ...828...85Y,2021MNRAS.506L...1D,2024arXiv241108498L}. Moreover, we find that the precision of the $ \Omega_K$ obtained through the ANN method is inferior to that achieved by the GP method. However, the central values of the curvature parameter obtained by both methods are nearly identical.
In conclusion, the estimations on $ \Omega_K$ obtained through any data combinations and  the reconstructed methods consistently support a flat universe within the 1$\sigma$ uncertainty. This conclusion differs from the recent work by \citet{2023PhRvD.108f3522Q}, as their study utilized SNe Ia data and relied on $H_0$ or absolute magnitude. This again highlights the importance of our approach.

By comparing and analyzing the samples using the two BAO data sources, we find that there is almost no difference between the two samples. Although the curvature values obtained from the data samples using DESI are on the slightly positive and the samples using eBOSS are on the slightly  negative, these results both report that our universe has a flat spatial curvature within uncertainties, and the precision of constraining the curvature is almost equal.

In the future, we will have access to vast amounts of BAO data, such as that from the ongoing DESI survey.   At the same time, we also expect that following surveys such as the WiggleZ Dark Energy Survey \cite{2010MNRAS.401.1429D}, future observations of cosmic chronometers will be greatly improved.
It is reasonable to expect that our method will play an increasingly important role in significantly improving curvature measurements with higher precision and accuracy using our method in the absences of cosmological model assumptions and cosmological parameters. However, using data reconstruction methods is still an effective technique and tool in the current data scarcity, and the rapid development and technological advancement of these methods likewise makes us optimistic about measuring curvature in the future with greater precision and accuracy.

\acknowledgments

This work was supported by National Natural Science Foundation of China under Grant No. 12203009, 12475051, 12122504 and No. 12035005; Hubei Province Foreign Expert Project (2023DJC040); The innovative research group of Hunan Province under Grant No. 2024JJ1006; the Natural Science Foundation of Hunan Province under grant No. 2023JJ30384; and the Hunan Provincial Major Sci-Tech Program under grant No.2023ZJ1010.

\end{document}